\documentstyle[aps,pra,epsf,twocolumn]{revtex}

\begin{document}
\draft

\tighten

\wideabs{

\title{Analytical description of a trapped semi-ideal Bose gas at finite
temperature}
\author{M.~Naraschewski}
\address{Jefferson Laboratory, Department of Physics,
Harvard University, Cambridge, MA~02138, USA}
\address{ITAMP, Harvard-Smithsonian Center for Astrophysics, 60 Garden Street,
Cambridge, MA~02138, USA}
\author{D.M.~Stamper-Kurn}
\address{Department of Physics and Research Laboratory of Electronics,
Massachusetts Institute of Technology, Cambridge, MA 02139}

\date{17 March 1998}

\maketitle

\begin{abstract}
Present experiments with Bose condensed gases can be largely described by a
semi-ideal two-gas model. In this model, the condensate is
influenced only by the mean-field repulsion among condensed atoms,
while the thermal cloud is considered an ideal gas confined by an effective
potential that consists of the external trap and the mean-field
repulsion by the condensate. This simple, intuitive model
provides explicit analytical expressions for the density distributions of
the condensate and the thermal component.
It describes the reduction of the condensate fraction
relative to that of an ideal gas as a consequence of the positive
chemical potential due to interactions in the condensate.
\end{abstract}

\pacs{03.75.Fi,05.30.Jp}

}

\narrowtext

Recent experiments with Bose-Einstein condensed alkali gases
(cf. e.g. \cite{anderson95,davis95b}) have been successfully described
by different approximation schemes. The diluteness of these gases,
expressed as $na^3\ll 1$ where $a$ is the two-body scattering length,
allows a description in terms of an effective mean-field
theory. At zero temperature the condensate can be described by the
Gross-Pitaevskii equation \cite{gross61,pitaevskii61}, a nonlinear
Schr\"odinger equation. In present experiments
interactions are strong enough so that the chemical
potential $\mu$ is much larger than the
level spacings $\hbar\omega_x$, $\hbar\omega_y$, and $\hbar\omega_z$ of the
external
harmonic potential. Therefore the use of the Thomas-Fermi approximation
\cite{goldman81,baym96}, which neglects the kinetic energy of the condensate, 
has lead to a tremendous simplification in the understanding of these
experiments.

Recently, much theoretical and experimental attention has been
focused on finite temperature properties of Bose-condensed gases.
A semiclassical
Hartree-Fock-Popov description has produced excellent agreement with
experimentally measured equilibrium quantities \cite{giorgini97a}.
Further, unlike in a spatially
homogeneous gas, it was found that collective elementary excitations
contribute very little to thermodynamic properties \cite{giorgini97a} due to
the relatively small spatial overlap between condensate and thermal component
in a harmonic trap. Thus, even a semiclassical Hartree-Fock
approximation is a fairly accurate description of these gases, except
for a small temperature range close to the critical temperature. For example,
excellent agreement has been achieved by a comparison of
Hartree-Fock density profiles with the results of a comprehensive
Path Integral Monte Carlo calculation \cite{holzmann98}.

In this paper we introduce a further simplification of the
self-consistent Hartree-Fock model by neglecting atomic interactions in the
thermal component of the gas. Like the introduction of the
Thomas-Fermi approximation \cite{goldman81} the reduction to Eqs.\ 
(\ref{n0model})--(\ref{mudef}) and (\ref{N0mf}) has the important
advantage of providing an explicit analytical description of a trapped Bose 
gas at finite temperature.

We consider atoms that are trapped in a generally anisotropic harmonic
potential. Under the appropriate rescaling of energy and lengths, the harmonic
potential has the form $V({\bf r}) = r^2/2$. Here, the
dimensionless cartesian coordinates of $\bf r$ are given in units of
$\sqrt{\hbar\omega/m\omega^2_{x,y,z}}$, with
$\omega=(\omega_x\omega_y\omega_z)^{1/3}$.
Other lengths, like the thermal wavelength
$\lambda_T$ and the scattering length $a$, are scaled
in units of the natural length $\sqrt{\hbar/m\omega}$ of a
harmonic oscillator with angular frequency $\omega$.
The energy $V({\bf r})$ as well
as other energies are implicitly given in units of $\hbar\omega$.

Then, in the self-consistent Hartree-Fock model, the densities of the
condensate and the thermal component are given
as solutions of the two coupled equations \cite{goldman81,giorgini97a}
\begin{eqnarray}
\label{n0stringari}
n_0({\bf r}) &=& \frac{\mu-r^2/2-2U n_T({\bf r})}{U}\,\theta
(\mu-r^2/2-2U n_T(r))\\
\label{nTstringari}
n_T({\bf r}) &=& \frac{1}{\lambda_T^3}
g_{3/2}\left(e^{-(r^2/2+2U\,[n_0({\bf r})+n_T({\bf r})]-\mu)/k_BT}
\right).
\end{eqnarray}
Here, the Thomas-Fermi approximation has been applied which neglects the
kinetic energy of the condensate and implicitly finite size effects.
The strength of the atomic interactions is given by $U=4\pi a$. We use
the thermal wavelength
$\lambda_T = \sqrt{2\pi/k_BT}$ and the
Bose function $g_{3/2}$ is defined by $g_\alpha(z)=\sum_{j=1}^\infty
z^j/j^\alpha$. The chemical potential $\mu$ is determined by
the constraint of a given total atom number
\begin{equation}
\label{Nfixed}
N=\int\!d{\bf r}\; \bigl[ n_0({\bf r}) + n_T({\bf r}) \bigr].
\end{equation}
For further reference, Eqs.\ (\ref{n0stringari}) --
(\ref{Nfixed}) will be denoted as the interacting model.
Even though it represents already a relatively concise description,
these equations still have to be solved self-consistently.
This is done numerically by an iterative procedure \cite{giorgini97a}.

If one further neglects the mean-field repulsion from non-condensed atoms
Eqs.\ (\ref{n0stringari}) and (\ref{nTstringari}) are solved by the explicit
relations
\begin{eqnarray}
\label{n0model}
n_0({\bf r}) &=& \frac{\mu-r^2/2}{U}\,\theta(\mu-r^2/2)\\
\label{nTmodel}
n_T({\bf r}) &=& \frac{1}{\lambda_T^3}
g_{3/2}\left(e^{-|r^2/2-\mu|/k_BT}\right),
\end{eqnarray}
with the chemical potential $\mu$ given by
\begin{equation}
\label{mudef}
\mu=\frac{1}{2}(15 N a)^{2/5}
\left(\frac{N_0}{N}\right)^{2/5}.
\end{equation}
These relations summarize a simple model of the partly condensed gas which we
denote as the semi-ideal model. The condensate density $n_0$ is that of a zero
temperature condensate with $N_0$ atoms, while
the density $n_T$ of the non-condensed cloud is that of an ideal
gas of bosons confined in the combination of the external potential
and the repulsive mean-field potential due to the condensate atoms.
Its wide range of validity has been confirmed by a numerical
analysis in Ref.\ \cite{dodd98}. However, in order to make it an
intuitively appealing, usable description of a trapped Bose gas, the
condensate fraction has to be determined analytically as a function of
temperature.

For later convenience let us introduce the reduced chemical
potential $\bar\mu$
\begin{equation}
\label{mubardef}
\bar\mu = \frac{\mu}{k_BT}=\eta\left(\frac{N_0}{N}\right)^{2/5}
\left(\frac{T}{T_c}\right)^{-1}.
\end{equation}
Here, the critical temperature $T_c$ is that of a trapped ideal gas
\cite{bagnato87,ketterle96a} $k_BT_c = [N/\zeta(3)]^{1/3}$
where $g_\alpha(1)=\zeta(\alpha)$ was expressed in terms of
the Riemann Zeta function.
The reduced chemical potential depends on the scaling parameter $\eta$
\cite{giorgini97a}
\begin{equation}
\label{etadef}
\eta = \frac{\mu_{T=0}}{k_BT_c} = \frac{1}{2}\,\zeta(3)^{1/3}
\left(15 N^{1/6} a\right)^{2/5}
\end{equation}
which describes the strength of the atomic interactions within the condensate.
The scaling parameter is independent of the system size when the thermodynamic
limit is taken
in the usual way ($N\to\infty,\omega\to 0,N\omega^3$ = const.). Due to its
relatively weak $N^{1/6}$ dependence, $\eta$ assumes a value close to
$0.3$ in most recent experiments \cite{giorgini97a}.

The condensate fraction is determined by integration over the thermal
density distribution. For $T<T_c$, the result can be written in terms of
incomplete Gamma functions \cite{gradshteyn80}
\begin{eqnarray}
\frac{N_0}{N} &=& 1 -
\frac{2}{\sqrt{\pi}}\frac{1}{\zeta(3)}\left(\frac{T}{T_c}\right)^{3}\nonumber\\
&&\times\sum_{j=1}^{\infty}\frac{1}
{(jk_BT)^{3/2}}\left(\int_0^\mu d\epsilon\sqrt{\epsilon}\,
e^{j(\epsilon-\mu)/k_BT}\right.\nonumber\\
&&\qquad\qquad\qquad + \left.\int_\mu^\infty d\epsilon\sqrt{\epsilon}\,e^{-j
(\epsilon-\mu)/k_BT}\right)\nonumber\\
&=& 1 - \frac{2}{\sqrt{\pi}}\frac{1}{\zeta(3)}\left(\frac{T}{T_c}\right)^{3}
 \sum_{j=1}^{\infty}\frac{1}{j^3}\,\left[e^{-j\bar\mu}\gamma(3/2,-j\bar\mu)
\right.\nonumber\\
\label{N0gamma}
&&\qquad\qquad\qquad + \left.e^{j\bar\mu}\Gamma(3/2,j\bar\mu)\right].
\end{eqnarray}
Eq.\ (\ref{N0gamma}) is still an implicit
expression since it depends again on the reduced chemical potential. Therefore,
$\bar\mu$ and $N_0/N$ are
given as simultaneous solutions of Eqs.\ (\ref{mubardef}) and
(\ref{N0gamma}).
For temperatures $T>T_c$ a condensate does not exist and the reduced
chemical potential $\bar\mu<0$ is trivially defined by the single equation
$g_3(e^{\bar\mu})=\zeta(3)(T/T_c)^{-3}$.

Let us consider the results of the semi-ideal model in the limits
of small and large $\bar\mu$, corresponding to
$T\approx T_c$ and $T\approx 0$. Eq.\ (\ref{N0gamma}) is therefore rewritten
\begin{equation}
\label{N0f}
\frac{N_0}{N} = 1 - \frac{2}{\sqrt{\pi}}\frac{1}{\zeta(3)}
\left(\frac{T}{T_c}\right)^{3}\sum_{j=1}^{\infty}\frac{1}{j^3}\,f(j\bar\mu)
\end{equation}
in terms of the function
\begin{equation}
\label{fdef}
f(x) = \int_0^{x} d\bar\epsilon\sqrt{\bar\epsilon}\,
e^{\bar\epsilon-x}
+ \int_{x}^\infty d\bar\epsilon\sqrt{\bar\epsilon}\,e^{-(\bar\epsilon-
x)},
\end{equation}
where the reduced energy $\bar\epsilon = \epsilon/k_BT$ has been introduced.
The function $f(j\bar\mu)$ can be expressed as a power
series in $\sqrt{j\bar\mu}$, where $\bar\mu$ runs from $0$ $(T=T_c)$ to
$\infty$ $(T=0)$. Truncating the series after the lowest nontrivial
order yields
\begin{eqnarray}
\frac{N_0}{N}&=&1-
\left(\frac{T}{T_c}\right)^3\left[1+\frac{\zeta(2)}{\zeta(3)}\bar\mu\right]
\nonumber\\
\label{N0mf}
&=&1-\left(\frac{T}{T_c}\right)^3-\eta\frac{\zeta(2)}{\zeta(3)}\left(\frac{T}
{T_c}\right)^2\left(\frac{N_0}{N}\right)^{2/5}
\end{eqnarray}
The first two terms of Eq.\ (\ref{N0mf}) correspond to
the condensate fraction of an ideal gas, whereas the third
describes the influence of the condensate repulsion to
lowest order in $\bar\mu$. This expansion adequately describes the solution of
Eq.\ (\ref{N0gamma}) over the entire range of temperatures, provided that
$\eta \ll 1$.
A further simplification is derived by solving Eq.\ (\ref{N0mf}) to
lowest order in $\eta$, arriving at
\begin{equation}
\label{N0explicit}
\frac{N_0}{N} = 1-
\left(\frac{T}{T_c}\right)^3-\eta\frac{\zeta(2)}{\zeta(3)}\left(
\frac{T}{T_c}\right)^2\left[1-\left(\frac{T}{T_c}\right)^3\right]^{2/5}.
\end{equation}

Within Eqs.\ (\ref{N0mf}) and (\ref{N0explicit}), the only effect of
the condensate interactions is
to shift the chemical potential to positive values,
raising the bottom of the effective potential
which the thermal atoms experience. The effect of changing the shape of the
potential enters only at higher orders of the expansion. Eq.\ (\ref{N0mf}) is
thus a good approximation of Eq.\ (\ref{N0gamma}) if most of the thermal
atoms are spatially separated from the condensate, a
condition satisfied for temperatures close to $T_c$ and for small $\eta$.

Eq.\ (\ref{N0mf}) is not
expected to be a good description for extremely low temperatures.
The understanding of this limit is facilitated by $f(x)$ obeying the
differential equation
\begin{equation}
\label{fdiff2}
\frac{d^2}{dx^2}f=f-2\sqrt{x}.
\end{equation}
The function $f(x)$ has to grow slower than $x^{3/2}$ for large
values of $x$. Otherwise, the power series in Eq.\ (\ref{N0f}) would not
converge. As a consequence, the second derivative in Eq.\
(\ref{fdiff2}) vanishes for large $x$. Thus, the asymptotic
limit of the condensate fraction for low temperatures is given by
$f(x) = 2\sqrt{x}$ or
\begin{equation}
\label{N0asymptotic}
\frac{N_0}{N}=1-
\sqrt{\eta}\frac{4}{\sqrt{\pi}}\frac{\zeta(5/2)}{\zeta(3)}\left(\frac{T}
{T_c}\right)^{5/2}.
\end{equation}
However, it has been shown \cite{giorgini97a} that the leading term of the
uncondensed fraction around zero temperature scales like
\begin{equation}
\label{N0phonon}
\frac{N_0}{N}=1-\eta\frac{\pi^2}{2\sqrt{2}\,\zeta(3)}
\left(\frac{T}{T_c}\right)^2
\end{equation}
due to quasi-particle contributions which are included in neither the
semi-ideal nor the interacting Hartree-Fock treatment.

\begin{figure}
\begin{center}
\leavevmode
\epsfxsize=0.45\textwidth
\epsffile{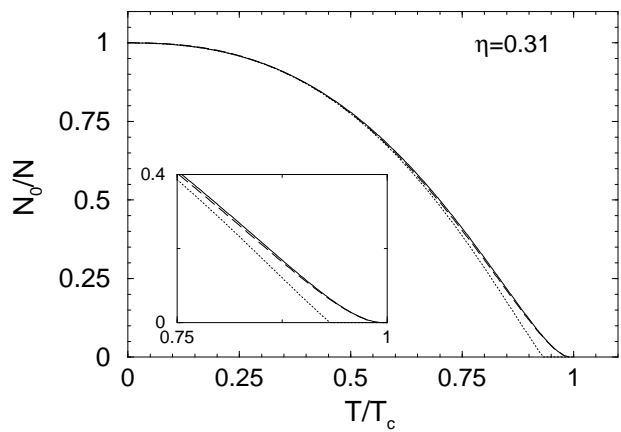}
\end{center}
\caption{\label{n0bfig}Condensate fraction as a function of temperature, given
by the semi-ideal model. The dashed line shows the exact solution 
Eq.\ (\ref{N0gamma}) of the semi-ideal model. The solid line represents
its high-temperature limit Eq. (\ref{N0mf}). The 
explicit expression of Eq.\ (\ref{N0explicit}) is given by the dotted line. 
This plot demonstrates the validity of Eq. (\ref{N0mf}) over the whole range of
temperatures $T<T_c$. The explicit approximation of Eq.\ (\ref{N0explicit}) 
fails only in the vicinity of the critical temperature (shown in the inset).
}
\end{figure}

According to Fig.\ \ref{n0bfig}, where $\eta = 0.31$ has been assumed,
the high temperature expansions Eqs.\ (\ref{N0mf}) and
to some extent Eq.\ (\ref{N0explicit}) are very good approximations of
the exact condensate fraction Eq.\ (\ref{N0gamma}) of the semi-ideal model 
over the whole range of temperatures $T<T_c$. In contrast, we have found that
the validity of the low temperature limits of Eqs.\ (\ref{N0asymptotic}) and
(\ref{N0phonon}) are restricted to a very small range around zero temperature
which makes them nearly irrelevant for practical purposes.

\begin{figure}
\begin{center}
\leavevmode
\epsfxsize=0.45\textwidth
\epsffile{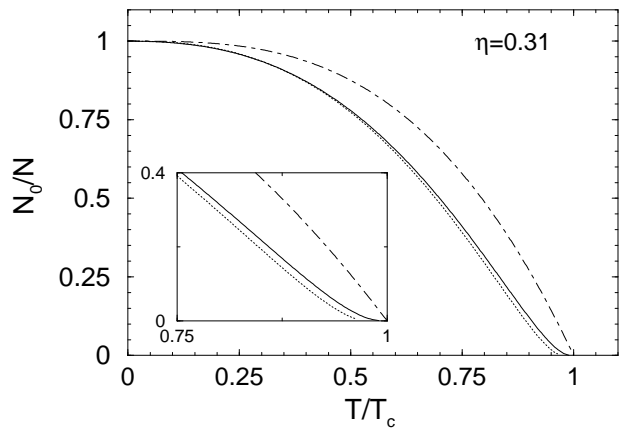}
\end{center}
\caption{\label{n0afig}Condensate fraction plotted as a function of
temperature, comparing the discussed models. 
The ideal gas condensate fraction is given by the dashed-dotted line.
The dotted line is a numerical
solution of the Hartree-Fock Eqs.\ (\ref{n0stringari})--(\ref{Nfixed}).
The solid line is obtained from the analytical approximation 
Eq.\ (\ref{N0mf}), corresponding to the solid line of Fig.\ \ref{n0bfig}.
Except near the critical temperature (shown in the inset), the semi-ideal 
model well approximates the solution of the interacting model.  
The critical temperature $T_c$ refers to the case of an ideal gas.}
\end{figure}

The condensate fraction given by the approximate solution of the 
semi-ideal model in Eq.\ (\ref{N0mf}) is 
compared in Fig.\ \ref{n0afig} with the result of an iterative solution of the
interacting model (Eqs.\ (\ref{n0stringari})--(\ref{Nfixed})).
The difference between the curves is small except near the critical
temperature, where the interactions among thermal atoms shift the
critical temperature \cite{giorgini96}. This shift is not exhibited by the
semi-ideal model. The wide range of validity of the
semi-ideal model has also been confirmed by a numerical comparison with a
Hartree-Fock-Popov calculation \cite{dodd98}. The latter includes collective
excitations, which are ignored in the models shown in Fig.\ \ref{n0afig}.
These lead to an additional but negligible decrease of the condensate fraction
\cite{giorgini97a}. 

\begin{figure}
\begin{center}
\leavevmode
\epsfxsize=0.45\textwidth
\epsffile{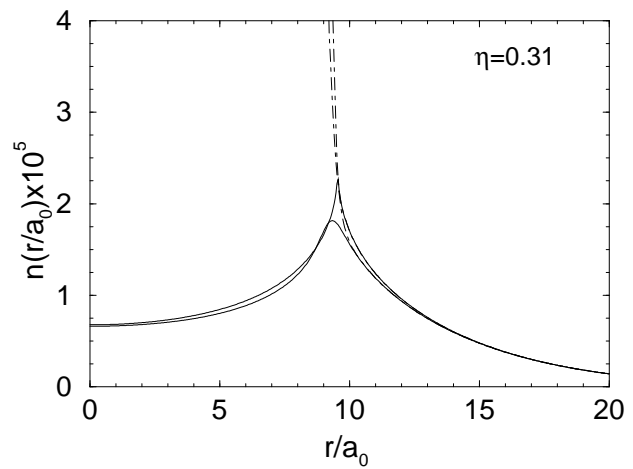}
\end{center}
\caption{\label{niafig}A comparison of the density distributions at
$T/T_c=0.5$ predicted by the two discussed models.
The upper solid line shows the density of the thermal
component as given by Eq.\ (\ref{nTmodel}). In comparison, the
lower solid line has been calculated numerically using Eq.\
(\ref{nTstringari}).
However, a numerical solution
of the full Gross-Pitaevskii equation has been used instead of the
Thomas-Fermi approximation of Eq.\ (\ref{n0stringari}). A small difference
exists at the edge of the condensate due to the neglect of the kinetic energy
in the semi-ideal model. However, the total density is only weakly affected
by this effect, as is seen by the two dashed-dotted lines. Again, the upper
dashed-dotted line refers to the total density given by the semi-ideal model
while the lower curve corresponds to the self-consistent Hartree-Fock 
approximation.
Here, a total atom number of $N=5\times 10^6$ was assumed.}
\end{figure}

A direct comparison between the density distributions of the
different models is given in Fig. \ref{niafig} for $T/T_c = 0.5$. Here,
the Thomas-Fermi solution of the condensate
density of Eq.\ (\ref{n0stringari}) has been replaced by a solution of the
full Gross-Pitaevskii equation. The main implication of the included kinetic
energy of the condensate is a smoothed out condensate surface.
Within the Thomas-Fermi approximation almost no
difference is seen between the two curves. Even though the unaccounted
existence of collective excitations changes the densities of condensate and
thermal component close to the center of the trap by a significant amount,
their influence on the total density is negligibly small
for most temperatures \cite{holzmann98}.

In recent experiments, trapped Bose gases have been probed in-situ
using either non-destructive dispersive
imaging techniques \cite{andrews96}, or off-resonant
absorption imaging \cite{hau98a}. These methods are important new tools in the
understanding of trapped Bose gases, with distinct advantages over the
previously used time of
flight measurements of expanding atom clouds \cite{anderson95,davis95b}.
We therefore expect that the results of this paper will become useful for
future in-situ experiments with trapped Bose gases.
These new optical detection techniques measure
column or line density profiles; we therefore determine these profiles
from Eqs.\ (\ref{n0model}) and (\ref{nTmodel}) by integration over one or
two dimensions.

In cylindrical coordinates, where the probe beam is directed along the
$z$-axis, the according column densities are given by
\begin{eqnarray}
n_0(\rho) &=&  \frac{4 \sqrt{2}}{3} \frac{(\mu-\rho^2/2)^{3/2}}{U}\,
\theta(\mu-\rho^2/2)\\
n_T(\rho) &=&  \frac{2\pi}{\lambda_T^4}
\left\{
\begin{array}{ll}
g_2(e^{\bar\mu(\rho)}) & \bar\mu(\rho)<0 \\ & \\
 \int_{0}^{\infty} d\bar\epsilon\,\frac{1}{\sqrt{\pi\bar\epsilon}}
g_{3/2}(e^{-|\bar\epsilon-\bar\mu(\rho)|}) & \bar\mu(\rho)>0
\end{array}
\right.
\end{eqnarray}
where $\bar\mu(\rho) = (\mu - \rho^2/2)/k_BT$.
Equivalently, we obtain the line densities
\begin{eqnarray}
n_0(z) &=& \pi \frac{(\mu-z^2/2)^2}{U}\,\theta(\mu-z^2/2)\\
n_T(z) &=& \frac{(2\pi)^2}{\lambda_T^5} \left\{
\begin{array}{ll}
g_{5/2}(e^{\bar\mu(z)}) & \bar\mu(z)<0 \\
2\zeta(5/2) - g_{5/2}(e^{-\bar\mu(z)}) & \bar\mu(z)>0
\end{array}
\right.
\end{eqnarray}
with $\bar\mu(z) = (\mu - z^2/2)/k_BT$.

We therefore conclude that the proposed semi-ideal model of Eqs.\
(\ref{n0model})--(\ref{mudef}) and (\ref{N0mf}) represents
a rather good description of a trapped Bose gas at finite
temperature provided that $\eta$ is not larger than
the values $\eta\approx 0.3$ to $0.4$ achieved in recent experiments.
Its mathematical simplicity gives a much clearer conceptual
picture than the self-consistent approaches that have been used before.
The observed agreement of the semi-ideal model with a
self-consistent Hartree-Fock calculation has confirmed
that the density of the thermal component is too low to necessitate an
interacting gas description. Instead, the observed depletion of the condensate
fraction arises from condensate interactions, leading to a positive value 
of the chemical potential or equivalently to a raised bottom of the 
effective potential in which the thermal component resides. It
is well described by Eq.\ (\ref{N0mf}), which is a
low order expansion in $\bar\mu$. 
In contrast, the changed shape of the effective
potential does not influence the thermal population in a noticeable way.

This work was supported by the Deutsche Forschungsgemeinschaft (DFG) and 
also the Office of Naval Research, the Joint Services Electronics Program
(ARO) and the JSEP Graduate Fellowship Program. It is
a pleasure to thank M. Olshanii, H.-J. Miesner, and W. Ketterle for many
valuable discussions. After the completion of this work, E. Mueller brought
to our attention a numerical calculation of the condensate fraction 
within the semi-ideal model \cite{minguzzi97}.

%\bibliographystyle{prsty}
%\bibliography{refbec,refman}

%\narrowtext
%\twocolumn

% Figures

\end{document}